# Dispersion Scan Frequency Resolved Optical Gating For Evaluation of Pulse Chirp Variation

**M. GUESMI,**[1,*] **PETRA VESELÁ,**[1] **AND KAREL ŽÍDEK**[1,*]

[1] *Regional Center for Special Optics and Optoelectronic Systems (TOPTEC), Institute of Plasma Physics, Czech Academy of Science v.v.i., Za Slovankou 1782/3, 182 00 Prague 8, Czech Republic*
*\*guesmi@ipp.cas.cz, zidek@ipp.cas.cz*

**Abstract:** The commonly used methods to characterize ultrafast laser pulses, such as frequency-resolved optical gating (FROG) and dispersion scan (d-scan), face problems when they are used on pulses with a chirp varying within the laser beam or the acquisition time. The presence of such chirp variation can be identified by a discrepancy between the measured FROG and d-scan traces and their reconstructed counterparts. Nevertheless, quantification of the variation from the experimental data is a more complex task. In this work, we examine the quantification of chirp variation based on three different pulse characterization techniques. Two commonly used techniques FROG and d-scan are compared to a new method dispersion scan FROG (D-FROG) that combines the idea of dispersion scanning with the FROG method. By using the three techniques, we analyze the chirp variation of pulses generated from NOPA together with pulses processed by a 4f-pulse shaper without and with SLM-adjusted phase. We evaluate the performance of the new D-FROG method for the chirp variation estimate and the improved reconstruction of the measured results. Furthermore, we discuss the origin of chirp variation in each of the measurements by using fast-scan autocorrelation traces.

## 1. Introduction

Ultrafast laser pulses are used in many different application domains, including the pulsed laser deposition [1], femtochemistry, femtobiology [2], or optical code division multiple access [3]. All the listed domains depend on reliable control and characterization of complex fs pulses, which can be carried out by a variety of methods. The simplest approach based on the autocorrelation technique provides only the minimum information about the shape of the pulse and a general waveform cannot be reliably recovered [4]. However, a range of other techniques makes it possible to recover even the complex pulse shapes. The most commonly used techniques are frequency-resolved optical gating (FROG) [5-7], cross-correlation frequency-resolved optical gating (XFROG) [6,7], spectral phase interferometry for direct electric-field reconstruction (SPIDER) [8], multiphoton intrapulse interference phase scan (MIIPS) [9], and dispersion scan (d-scan) [10].

Many of these methods face a problem when they are applied to characterize a pulse shape varying the laser beam or the acquisition time. We denote this situation as a chirp variation. In other words, the majority of standard pulse retrieval methods, including FROG, SPIDER, and d-scan, assume that the pulse shape is constant throughout the measured beam. However, even a small misalignment of a pulse compressor or a pulse shaper introduces a certain amount of spatial chirp (also denoted as chirp distortion) [11-13]. The spatially distorted pulse chirp also occurs due to the space-time coupling, which inevitably arises in the 4f pulse shapers [14,15]. An analogous situation can occur when the measured pulse is not stable in time and its chirp rapidly varies during the measurement. An example of this can be the effect of thermal fluctuations in spatial light modulators (SLMs) [16]. The chirp variation manifests itself in the standard FROG setup as a discrepancy between the measured and the retrieved FROG traces. Nevertheless, the quantification of distortion from the FROG experimental data represents a more challenging task that has not been yet addressed.

In this article, we study the use of three different methods to quantify the chirp variation. Two commonly used techniques, FROG and d-scan, are compared to a new method, where we acquire FROG traces while we controllably vary the pulse dispersion by a known value of group delay dispersion (GDD) and third-order dispersion (TOD). We denote this method as dispersion scan FROG, shortly D-FROG. The D-FROG dataset includes both the d-scan and a range of FROG traces. We demonstrate the use of the three different methods on a characterization of pulses (i) generated from a non-collinear optical parametric amplifier (NOPA); (ii) the NOPA pulse passing through a 4f-pulse shaper without any modulation device; (iii), the NOPA pulse passing through a 4f-pulse shaper with SLM-adjusted phase.

We fit the experimental data by a model, where we assume that the GDD values are not a single value, which would correspond to an ideal pulse, but rather a distribution of the GDD values. The width of the distribution reflects the pulse variation. To reveal the origin of this chirp variation, we used a rapid autocorrelation scan, which allowed us to assign the variation to the chirp distortion across the laser beam.

By scrutinizing the attained results, we show that the D-FROG method provides a means of careful evaluation of the laser pulse, which can reveal and quantify the variation of the pulse chirp on the order of tens of $fs^2$. We also demonstrate that even in the cases, where a single measured FROG trace can be well reproduced with an ideal pulse, the D-FROG data provide a consistency checkable to identify the chirp variation.

## 2. Experimental procedure

The used experimental setup is depicted in Fig. 1. We employed a fs laser system Pharos (Light Conversion) operated at 1028 nm, which generated pulses 290 fs long at 10 kHz repetition rate, 100 µJ/pulse. A part of the output power (50 µJ/pulse) was converted by a non-collinear optical parametric amplifier (NOPA) N-3H Orpheus (Light Conversion) into a visible laser pulse at 640 nm, spectral width FWHM 757 $cm^{-1}$. We used a prism pulse compressor integrated into the NOPA to adjust pulse length and to vary the dispersion. The prism consisted of fused silica prisms separated by 775 mm, where the prism P2 insertion was adjusted by a motorized stage to alter the pulse dispersion. It is worth noting that the measured pulse did not represent the best attainable pulse compression of the NOPA pulses. The presence of distinct features in the FROG traces was beneficial for the demonstration of the instability effect and dispersion scanning.

The pulse from NOPA was either directly characterized in a FROG setup, which is described later, or it was modified by a pulse shaper. We used a standard 4f pulse shaper – see Fig. 1b -- employing grating 600 gr/mm, which spectrally disperse the beam in the horizontal direction. The dispersed beam was collimated and focused in the vertical direction by a spherical mirror ($f$ = 500 mm) onto a spatial light modulator SLM-S640 (Jenoptik) placed in the Fourier plane. SLM was calibrated by using a procedure described in [17]. Consequently, symmetrically aligned mirrors were used to refocus the beam on the grating.

Finally, the temporal and spectral shape of the output pulse was measured by using the FROG setup. In this setup, the pulse was split with a pair of 50:50 beam splitters into two pulse replicas with an identical chirp. The delay between the pulses was varied by a motorized delay line (PIMag Linear Stage) and the pulses were focused on a 0.05 mm thick beta barium borate (BBO) crystal (Eksma Optics), where they generated a sum-frequency signal. Due to the non-collinearity of the incident pulses, the sum-frequency signal can be spatially separated by a pinhole and coupled with a lens into a fiber and analyzed by a spectrometer (AvaSpec-ULS4096CL-EVO). The laser spectrum was measured from a scattered beam on the pinholes (Ocean Optics Flame T).

The second harmonic generation autocorrelation (SHG-AC) was acquired for the rapid scanning as a sum of the measured spectra. The acquisition time of a single SHG-AC trace was 60 ms. The traces were measured in the sweep mode, ie. continuously moving delay line with velocity 1 mm/s, spectrometer integration time 1 ms. To attain a sufficiently high signal, we increased in this experiment the repetition rate of the laser to 100 kHz.

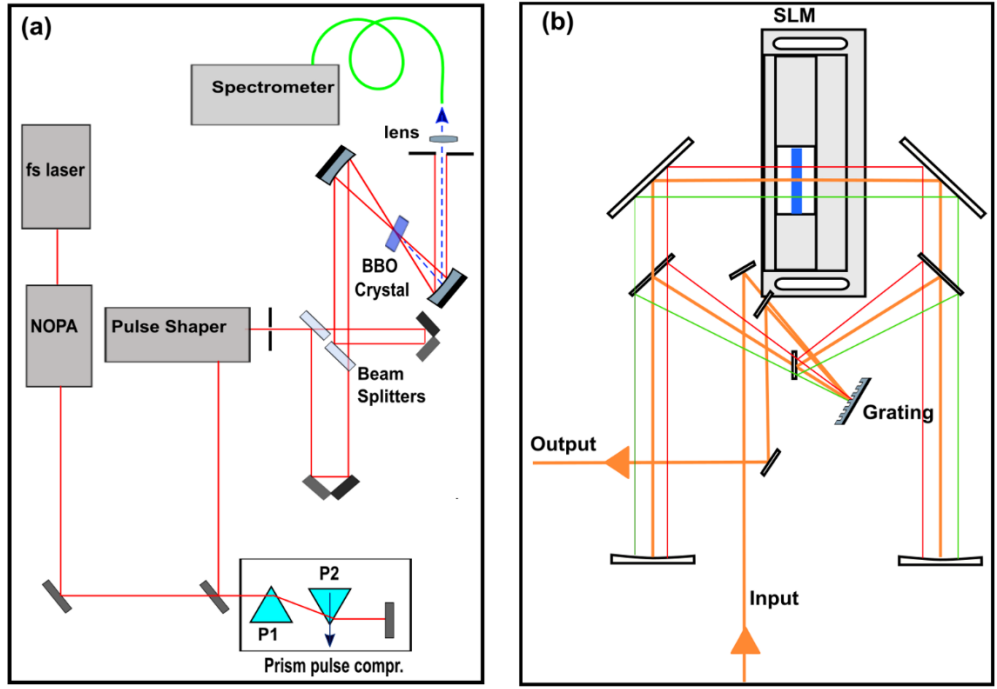

Fig. 1: (a) Experimental setup for the D-FROG. (b) Pulse shaper in 4f geometry.

## 3. Methods

### 3.1 FROG

Our setup is based on the second harmonic FROG, where the intensity of the measured trace is given by:

$$I_{FROG}(\omega,\tau) = |\int E(t)E(t-\tau)\exp(-i\omega t)dt|^2, \qquad (1)$$

where $\tau$ and $\omega$ are the pulse delay and the light frequency, respectively. $E(t) = \sqrt{I(t)}\exp(-i\Phi(t))$ is a complex amplitude of the pulse, where $I(t)$ is proportional to the light intensity and $\Phi(t)$ denotes the phase as a function of time. The spectrum for $\tau = 0$ corresponds to the SHG from a single pulse used in the d-scan measurements:

$$I_{SHG}(\omega) = |\int E(t)^2\exp(-i\omega t)dt|^2. \qquad (2)$$

A wide variety of algorithms can be used to retrieve the original pulse field $E(t)$ from the from trace $I_{FROG}(\omega,\tau)$. We employed a ptychographic reconstruction procedure by Sidorenko et al. [18], which is illustrated in Fig. 2. Before the reconstruction, the experimental SHG FROG trace (upper left panel) was interpolated from the measured data to form $N \times N$ matrix to reconstruct $N$ elements of intensity and phase vector. The depicted experimental trace represents the electric field amplitude, i.e. $\sqrt{|I_{FROG}(\omega,\tau)|}$. This was very well reproduced by the theoretical trace (upper right panel). The pulse reconstruction provided us with the temporal and spectral intensities and phases of the pulse depicted in the lower panels.

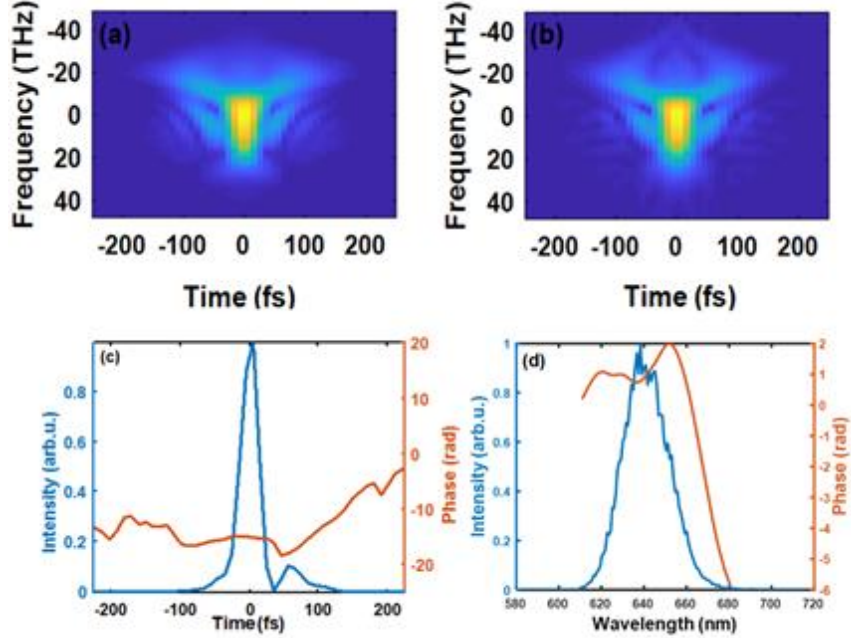

Fig. 2: Measured (a) and Retrieved (b) FROG traces from a compressed NOPA pulse. (c) Retrieved temporal intensity profiles and phases. (d) Retrieved spectral intensity profiles and phases. N=156, T=800, dt=10.32 fs.

In all the presented reconstructions, we corrected the FROG trace for marginals, which nevertheless lead only to subtle changes in the FROG trace owing to the use of the very thin SHG-generating crystal (0.05 mm). We also applied the measured laser spectrum in the reconstruction. Starting from the 11$^{th}$ iteration, we replaced the retrieved spectrum of the pulse amplitude with the measured one. The FROG trace reconstruction was used as an initial step in all the presented measurements to estimate the spectral phase of the pulse $\varphi_C$ for the reference prism insertion $\Delta p = 0$.

The FROG reconstruction, where we extract $2N$ values from $N^2$ is an overdetermined system, which can be used to check the consistency of the results. Inconsistency leads to a difference between the experimentally measured $I_{FROG}^{exp}$ and the theoretically retrieved $I_{FROG}^{th}$ FROG traces, which is commonly evaluated by the G-error:

$$G_{FROG} = \sqrt{\frac{1}{N^2} \sum_{\omega,\tau} \left| (I_{FROG}^{exp}(\omega, \tau) - \mu \cdot I_{FROG}^{th}(\omega_i, \tau)) \right|^2}, \quad (3)$$

where both experimental and theoretical traces are normalized to their maximum value and the parameter $\mu$ is optimized to obtain the minimum $G$ value. A high value of the error $G$ indicates an issue in the measurement. However, it is problematic to judge the actual origin of the issue only from the experimental FROG trace, because the increased G error is affected by many factors, including the noise level of the FROG spectra [19], delay line imprecision [20], or chirp variation.

## 3.2 Dispersion-scan FROG

To overcome this problem, we employed an additional consistency check, where we controllably scan dispersion over a selected range to observe the phase change of the reconstructed pulse and the corresponding FROG trace. We scan the dispersion by varying position $p$ of a prism in the prism compressor (see experimental section). In the spectral domain, the electrical field is given by: $\tilde{E}(\omega) = \sqrt{S(\omega)} \exp(i\varphi(\omega)) = \mathcal{F}\{E(t)\}$, where $\sqrt{S(\omega)}$ and $\varphi(\omega)$ are the spectral amplitude and phase, respectively. The phase $\varphi(\omega, \Delta p)$ for a certain change in prism insertion $\Delta p$ is given by a Taylor series as:

$$\varphi(\omega, \Delta p) = \varphi_C + \varphi_1(\omega - \omega_0) + \frac{1}{2}GDD(\Delta p).(\omega - \omega_0)^2 + \frac{1}{3}TOD(\Delta p).(\omega - \omega_0)^3, \quad (4)$$

where $\omega_0$ is the central frequency, $GDD$ represents the group delay dispersion and $TOD$ is the 3rd-order dispersion. $\varphi_C$ stands for the phase of the pulse for $\Delta p = 0$. The second term $\varphi_1$ causes the pulse translation in time and it can be ignored in the FROG calculation. At the same time, the higher dispersion terms can be neglected.

## 3.3 Chirp variation

Under certain *conditions*, for instance, due to the space-time coupling induced in a 4fpulse shaper, it has been observed that pulse varies across the laser beam [14,15]. In this case, FROG trace becomes a sum of FROG traces with a distribution of spectral chirp $\varphi(\omega, \Delta p)$.
We can simulate the chirp variation by assuming that the resulting FROG trace arises as a sum of FROG traces from a set of pulses, where the quadratic chirp GDD is not a single value $GDD(\Delta p)$, but rather a distribution $D(g)$. We will use in the article a Gaussian distribution centered around the mean value $GDD(\Delta p)$ with the standard deviation $\sigma_{GDD}$:

$$D(g) = \frac{1}{\sqrt{2\pi\sigma_G^2}} \exp\left[-\frac{(g - GDD(\Delta p))^2}{2\sigma_{GDD}^2}\right]. \quad (5)$$

Nevertheless, a various set of models can be applied, depending on the expected chirp behavior. The field corresponding to the GDD value $g$ can be expressed as:

$$E(t, g, \Delta p) = \mathcal{F}^{-1}\left\{\sqrt{S(\omega)}.\exp\left[\varphi_C + \frac{1}{2}g.(\omega - \omega_0)^2 + \frac{1}{3}TOD(\Delta p).(\omega - \omega_0)^3\right]\right\}, \quad (6)$$

which allows us to calculate the D-FROG signal $I_{DF}$ from a distorted pulse as a function of prism insertion:

$$I_{DF}(\omega, \tau, \Delta p) =$$

$$\frac{1}{\sqrt{2\pi\sigma_{GDD}^2}} \int \exp\left[-\frac{(g - GDD(\Delta p))^2}{2\sigma_{GDD}^2}\right] |\int E(t, g, \Delta p) E(t - \tau, g, \Delta p) \exp(-i\omega t) dt|^2 dg. \quad (7)$$

By using Eq. (6), we can extract the d-scan signal $I_{DS}$ by setting $\tau = 0$, ie

$$I_{DS}(\omega, \Delta p) = \frac{1}{\sqrt{2\pi\sigma_{GDD}^2}} \int \exp\left[-\frac{(g - GDD(\Delta p))^2}{2\sigma_{GDD}^2}\right] |\int E(t, g, \Delta p)^2 \exp(-i\omega t) dt|^2 dg. \quad (8)$$

To theoretically calculate the D-FROG, FROG and d-scan signal, we need to determine the GDD and TOD change with prism insertion $GDD(\Delta p)$ and $TOD(\Delta p)$. These can be attained by using the approach introduced as a self-calibrating d-scan [22,23]. The spectral intensity shape $S(\omega)$ was determined based on the laser spectrum. Finally, the spectral phase $\varphi_C$ for the reference prism insertion $\Delta p = 0$ was estimated by a standard FROG trace reconstruction (described in detail later). Therefore, we could use Eqs. (7) and (8) to fit a single unknown parameter $\sigma_{GDD}$ from FROG, D-FROG and d-scan datasets. The quality of the fit for the FROG trace can be evaluated based on the *G*-error in Eq. (3). For the D-FROG and d-scan data, we introduced an analogous *G*-error definition, where we sum the data also over the set of prism insertions $\Delta p_i$ i:

$$G_{DF} = \sqrt{\frac{1}{\mathcal{N}_i N^2}\sum_{\omega,\tau,i}\left|(I_{DF}^{exp}(\omega,\tau,\Delta p_i) - \mu.I_{DF}^{th}(\omega,\tau,\Delta p_i))\right|^2} \,, \tag{9}$$

$$G_{DS} = \sqrt{\frac{1}{\mathcal{N}_i N^2}\sum_{\omega,\tau,i}\left|(I_{DS}^{exp}(\omega,\Delta p_i) - \mu.I_{DS}^{th}(\omega,\Delta p_i))\right|^2} \,. \tag{10}$$

In this case, we use an additional factor $\mathcal{N}_i$, which indicates the number of scanned prism insertion positions.

## 4. Results and discussion

We used three different methods: FROG, D-FROG, and d-scan to study and quantify the chirp variation. We will first introduce our methods on a simple case, where we directly measured a pulse generated from NOPA, which bypassed the pulse shaper.

The D-FROG and d-scan datasets were attained by scanning the prism insertion and FROG trace acquisition for each of the positions – see upper panels in Fig. 3 (a) – and d-scan trace depicted in Fig.3 (b) – upper panel. To retrieve the corresponding theoretical traces, we needed to find the dispersion change with the prism insertion, which we assumed to change linearly: $G(\Delta p) \propto \Delta p$ and $T(\Delta p) \propto \Delta p$. The values were extracted by using the self-calibration d-scan approach, where we fitted the d-scan experimental trace by using Eq. (2). We set the FROG-reconstructed pulse as an initial guess, while the $GDD(\Delta p)$ and $TOD(\Delta p)$ scaling was left as a free fitting parameter. We extracted the values of $GDD(\Delta p)/\Delta p$ = 500 fs$^2$/mm and $TOD(\Delta p)/\Delta p$ = 10 fs$^3$/mm.

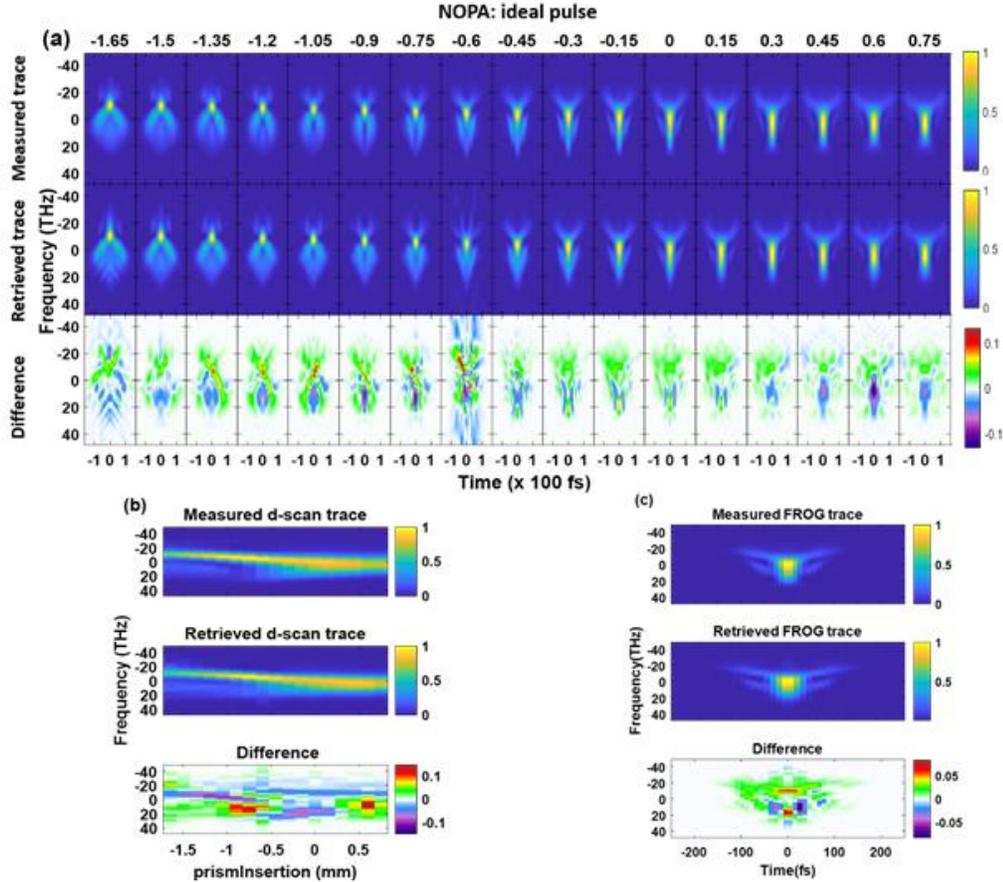

Fig. 3: Measured and retrieved traces of an ideal pulse, generated from NOPA, and their differences. (a) D-FROG traces (the traces are acquired for different prism insertion positions from -1.65 to 0.75 mm, as stated above the panels) (b) d-scan trace (c) FROG trace of a compressed pulse. The traces are retrieved using Eqs. (1) and (2), respectively, where *N=156, T=800, dt=10.32 fs*.

This allowed us to calculate the FROG, D-FROG and d-scan traces, which we first evaluated from Eqs. (1) and (2) for an ideal pulse. Even by assuming the ideal chirp stability, we attained for the FROG traces the *G* error of about 0.5%, which is a value commonly-obtained for the SHG FROG experiments and the remaining minor difference can arise, for instance, by an error in the delay line position, imperfections in the spectral calibration of the spectrometers, or other factors [19,20]. It is worth stressing that the agreement in 17 FROG traces in Fig. 3a with very different shapes is achieved by using a single reconstructed spectral phase $\varphi_C$ and two values of $GDD(\Delta p)$ and $TOD(\Delta p)$.

The small G errors between the measured and recovered traces for the NOPA pulse (listed in Table 1 for all experiments) are a sign of the low chirp variation. This is also confirmed by the d-scan trace in Fig. 3b, which is not smeared along the prism insertion axis, as it would be expected for the distorted chirp. We can therefore expect a subtle chirp variation in this dataset.

To study and quantify the chirp variation, we fitted the FROG, D-FROG and d-scan traces from Eqs. (7) and (8) by leaving the chirp distribution width $\sigma_{GDD}$ as a free fitting parameter. The retrieved D-FROG and FROG traces are shown in Fig. 4 (a) and Fig. 4 (c), where their fitted standard deviations $\sigma_G$ are equal to 30 and 0 fs$^2$, respectively. The difference between each measured and calculated trace is depicted in the panel below with a different color scheme.

Naturally, by using an additional fitting parameter, we were able to improve the retrieved traces and lowers the *G* error values (see Table1). This improvement is visually apparent for the lowest prism insertion $\Delta p = -1.65$ mm, where the fine structure present in the ideal reconstruction (see positive frequencies) is smeared by the chirp variation, and for the prism insertion $\Delta p = -0.6$ mm, where the same takes place for the fine structure at the negative frequencies. Nevertheless, in agreement with the qualitative analysis, the chirp variation is close to zero. Such small variation can arise due to a slight misalignment of the prism compressor or by the pulse generation in NOPA, where both can cause a minor spatial chirp variation across the beam.

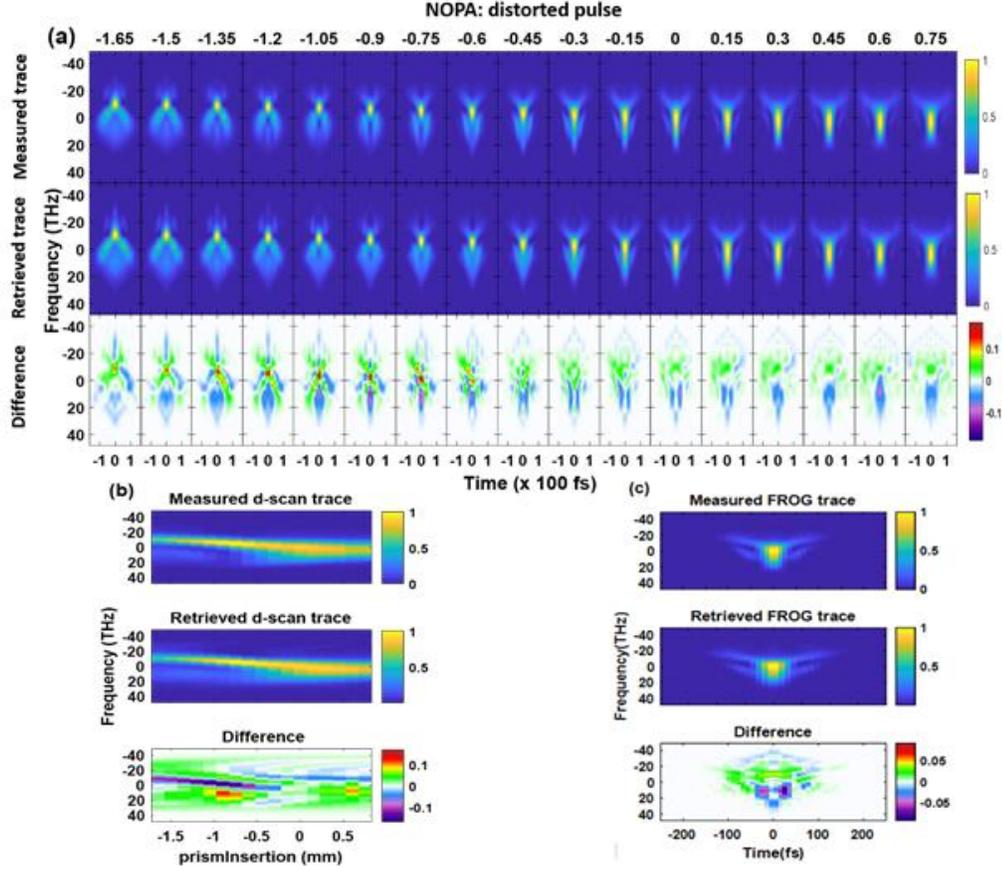

Fig. 4: Measured and retrieved traces of a distorted pulse, generated from NOPA, and their differences. (a) D-FROG traces (the traces are acquired for different prism insertion positions from -1.65 to 0.75 mm, as stated above the panels) (b) d-scan trace (c) FROG trace of a compressed pulse. The traces are retrieved using Eqs. (7) and (8), respectively, where *N=156, T=800, dt=10.32 fs*.

Since the pulses generated from NOPA are expected to feature very low chirp variation, we extended our methods on the characterization of pulses processed with a 4f-pulse shaper. Firstly, we evaluated the configuration, where the 4f-pulse shaper is used without the SLM device. The measured D-FROG, d-scan, and FROG trace are depicted in Fig. 5 and they are compared to the theoretical traces computed by using Eqs. (7) and (8). The spectral phase was extracted from the FROG experiment at $\Delta p = 0$, analogously to the previous dataset. In this case, the use of an ideal pulse without chirp variation leads to a significant difference between the measured and calculated traces—see Fig. S1 in the Supplementary material. This is also reflected by an increased G error of 0.79%, which becomes higher, while all experimental parameters, including the noise level, remained the same. A likely explanation is the presence of chirp variation.

To quantify the level of chirp variation, we fitted the D-FROG traces and observed that the fitted value of chirp variation, which reproduced best the experimental D-FROG data, reached 120 $fs^2$. In contrast to the NOPA pulses, here the inclusion of the chirp variation highly improved the G error to the level of the original NOPA pulses (0.79% before, 0.52% after). When we applied the minimization of the G error with respect to the chirp variation on the d-scan trace in Fig. 5 (b), we also reached the same value of 120 $fs^2$. By fitting the chirp variation by using a single FROG trace (at $\Delta p = 0$) in Fig. 5 (c), we derive the chirp variation of 80 $fs^2$.

While the D-FROG and d-scan data provided us with consistent values, the optimization of a single FROG trace provided a significantly lower value. This result can be explained by the fact that since the FROG trace is used for the phase reconstruction and the retrieved phase can partly compensate for the chirp variation. Therefore, we tested the optimization of $\sigma_{GDD}$ value on three different spectral phases $\varphi_C$, where two were extracted from the FROG data (ptychograhic reconstruction) and one derived from d-scan pulse retrieval, which followed the procedure of Ref. [23]. All three phases are compared in Fig. 6(a). Phase 1 (black line in Fig. 6(a)) corresponds to the spectral phase employed in the previous reconstruction. Note that the phases differ substantially only in the parts, where the laser spectrum is below 5% of the peak intensity, otherwise, the differences are rather minor.

We studied the $G$ error as a function of the chirp variation $\sigma_{GDD}$ for each of the characterization methods (FROG, d-scan and D-FROG) and spectral phase – see Fig. 6(b). Each panel in Fig. 6 (b) corresponds to one curve in Fig. 6(a). For the sake of better comparison, the $G$ error data were normalized on their lowest value.

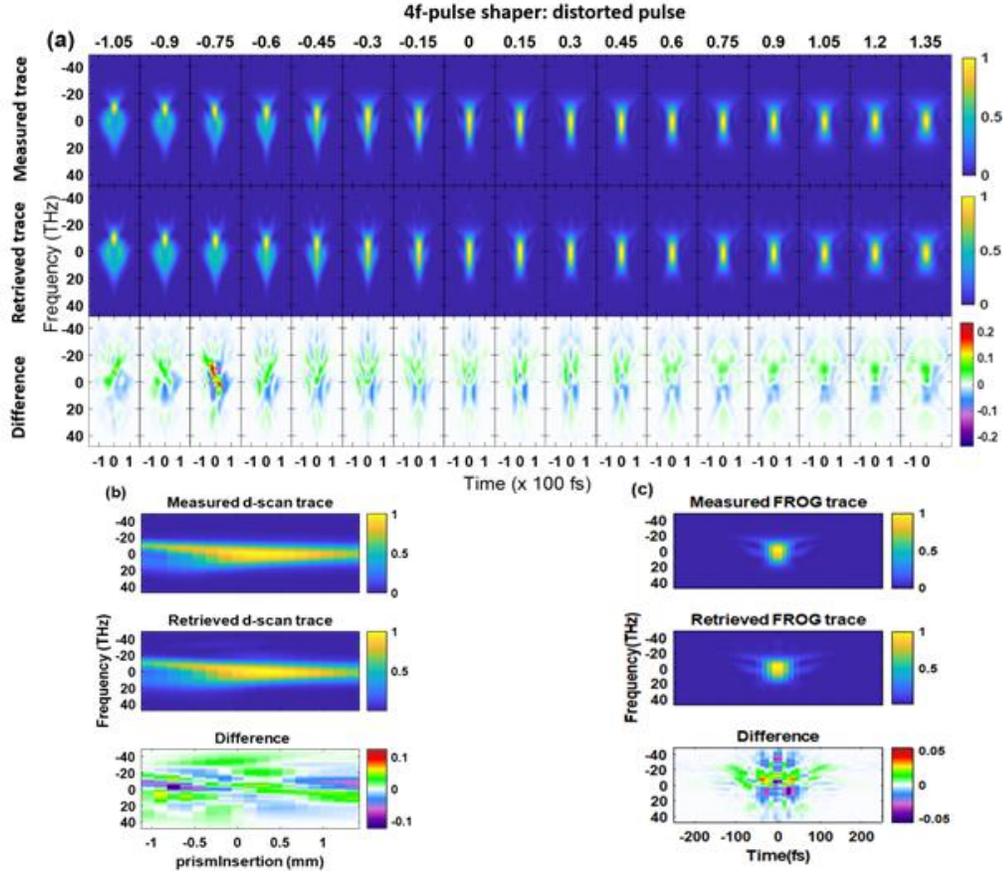

Fig. 5: Measured and retrieved traces of a distorted pulse, generated from NOPA and measured with a 4f-pulse shaper, and their differences. (a) D-FROG traces (the traces are acquired for different prism insertion positions from -1.05 to 1.35 mm, as stated above the panels) (b) d-scan trace (c) FROG trace of a compressed pulse. The traces are retrieved using Eqs. (7) and (8), respectively, where N=156, T=800, dt=10.32 fs.

Table 1. the $G$ error and chirp variation $\sigma_{GDD}$ values estimated using the FROG, D-FROG and d-scan methods of the pulses generated from a NOPA then measured through a 4f-pulse shaper (PS) without and with SLM-adjusted phase (PS+SLM). The ideal case corresponds to the value $\sigma_{GDD}=0$.

| Methods | FROG | | | D-FROG | | | d-scan | | |
|---|---|---|---|---|---|---|---|---|---|
| Pulses | ideal | distorted | | ideal | distorted | | ideal | distorted | |
| Values | $G$ (%) | $\sigma_{GDD}$ (fs$^2$) | $G$ (%) | $G$ (%) | $\sigma_{GDD}$ (fs$^2$) | $G$ (%) | $G$ (%) | $\sigma_{GDD}$ (fs$^2$) | $G$ (%) |
| NOPA | 0.49 | 0 | 0.49 | 0.66 | 30 | 0.64 | 0.31 | 30 | 0.34 |
| PS | 0.60 | 80 | 0.37 | 0.79 | 120 | 0.52 | 0.32 | 120 | 0.20 |
| PS+SLM | 0.72 | 120 | 0.57 | 0.81 | 180 | 0.57 | 0.36 | 180 | 0.25 |

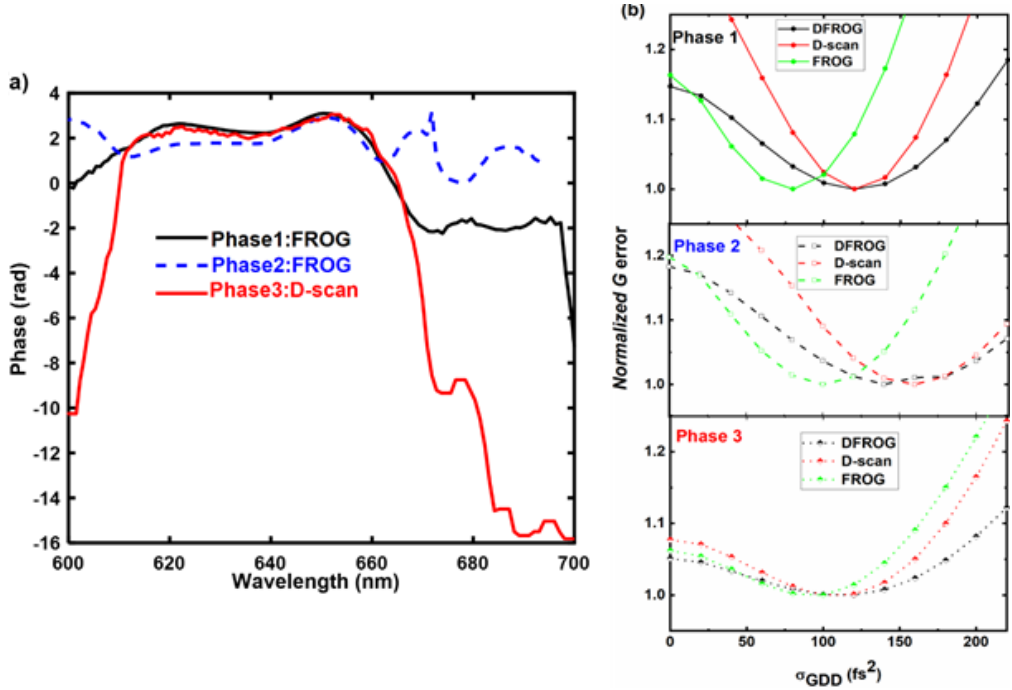

Fig. 6: (a) Spectral phases, retrieved using FROG and d-scan methods, of a pulse generated from a NOPA and measured through a 4f-pulse shaper without any modulation device. (b) Normalized $G$ error versus the chirp variation $\sigma_{GDD}$, using three different phases. We observe that depending on the initial spectral phase $\varphi_C$, the values can differ in the order of tens of fs$^2$.

To elucidate more the sensitivity of $\sigma_{GDD}$ value extraction with respect to the use of different phases, we carried out a set of calculations, where we extracted the pulse spectral phase from the FROG reconstruction initiated by randomly varied phases. For each reconstruction, we scanned the $\sigma_{GDD}$ value between 0 to 200 fs$^2$ and calculated the corresponding $G$ error for all three methods. This allowed us to extract the optimized $\sigma_{GDD}$ value, which varied for all the methods with the standard deviation of 10 fs$^2$.
While the FROG method (green lines in Fig. 6(b)) can reveal the chirp variation and provide a consistent value, it tends to underestimate the chirp variation. The d-scan method (red lines) is more reliable to detect the chirp variation than FROG. However, when we used the d-scan for the phase retrieval, the method also partly compensated for the chirp variation in the pulse shape (see Fig. 6(b), lowest panel). In other words, the pulse characterization by the d-scan or FROG only leads to an underestimated value of $\sigma_{GDD}$. Finally, the D-FROG method was the most robust one to extract and quantify chirp variation, where we observe differences of tens of fs$^2$ based on different approaches to reconstructions of the experimental data.

As the final step, we characterized pulses generated from a NOPA and processed by a 4f-pulse shaper with an SLM-adjusted phase. The SLM phase was adjusted to provide a constant phase, i.e. not to alter the pulse shape. Based on the measured DFROG, d-scan, and FROG traces compared with calculated ideal traces by using Eqs. (1) and (2), see Fig. S2 in the Supplementary material, we expect the presence of chirp variation and therefore we present the fit of the experimental data with Eqs. are (7) and (8) – see Fig. 7. By comparing this dataset to the D-FROG traces measured without the SLM (Fig. 5), we observed both D-FROG a d-scan experimental traces to be smeared along their horizontal axis, confirming the presence of the chirp variation.

The extracted chirp variation $\sigma_{GDD}$ increased by the SLM involvement from 120 to 180 fs$^2$, which was consistently obtained from the D-FROG and d-scan datasets. Analogously to the previous case, the inclusion of chirp variation decreased the *G*-error to the level induced by the experimental imperfections themselves, which was observed for the NOPA pulses.

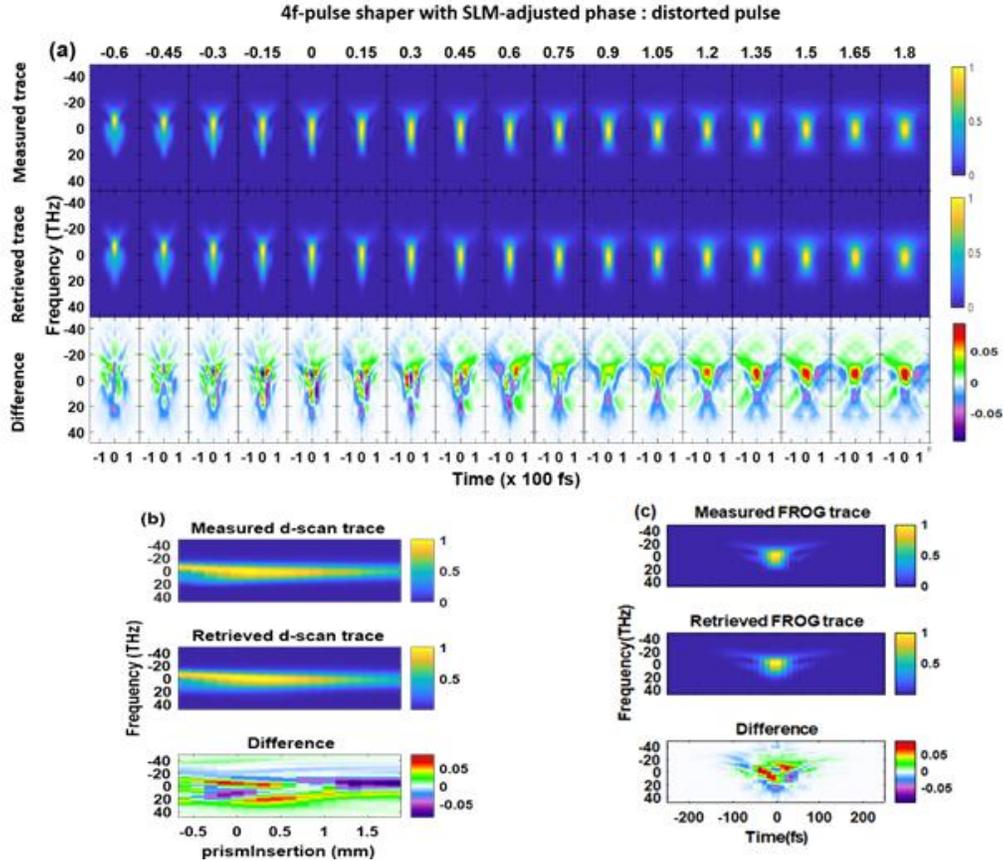

Fig. 7: Measured and retrieved traces of a distorted pulse, generated from NOPA and measured with a 4f-pulse shaper with SLM-adjusted phase, and their differences. (a) D-FROG traces (the traces are acquired for different prism insertion positions from -0.6 to 1.8 mm, as stated above the panels) (b) d-scan trace (c) FROG trace of a compressed pulse. The traces are retrieved using Eqs. (7) and (8), respectively, where *N=156, T=800, dt=10.32 fs*.

An important question of interest was the actual source of the chirp variation. The increased value of $\sigma_{GDD}$ can be both the signature of chirp instability in time, as well as chirp variation across the measured laser beam, often denoted as a spatial chirp or chirp distortion. To discriminate between the two cases, we measured a series of FROG traces by using a rapid delay line sweeping. We converted the traces by spectral integration into the second harmonic generation autocorrelation (SHG-AC) traces. Even for the acquisition time of 60 ms per scan (see Supplementary material for the data), all the measured cases show ideally stable autocorrelation traces. This means, that the observed chirp variations for the pulses processed by the pulse shaper originate very likely from the spatial chirp distortion. The occurrence of chirp temporal instability would need to take place on the timescale significantly below 60 ms, which is very unlikely in light of the previous results from the group of M. Motzkus, where the vast majority of noise level on a similar experimental configuration was present on frequencies < 20 Hz (timescale > 50 ms) [16]. Hence, we can conclude that the dominating origin of the chirp variation is the spatial chirp distortion across the laser beam. Such distortion can arise due to the misalignment of the PS [21], which is even more pronounced for the SLM by the pixelation in the SLM and its position outside of the Fourier plane [14,15].

## 5. Conclusion

In conclusion, we present a method of dispersion-scan frequency-resolved optical gating (D-FROG), which measures FROG traces as a function of dispersion and combines the idea of the FROG technique and the d-scan. We used this method to quantify the chirp variation of the ultrafast pulse.

By employing a model that assumes that the quadratic chirp GDD is not a single value but rather a Gaussian distribution, we studied the ability of the FROG, D-FROG and d-scan methods to quantify the chirp variation. We applied the methods on the characterization of pulses generated from a NOPA and the same pulses processed by a 4f-pulse shaper, which was used with and without SLM. By extracting the chirp variation for different pulse retrieval approaches, we observed that the D-FROG method provided the most consistent values of chirp variation. This can be understood in terms of the large dataset, which provides a broad consistency check.

Our measurement showed that while the NOPA-generated pulses are ideal throughout the measurement, the chirp variation appears for the pulses processed by the pulse shaper (approx. 120 $fs^2$) and it becomes larger with the use of SLM (180 $fs^2$). Nevertheless, the rapid-scan measurements, where a single FROG trace was acquired within 60 ms, did not show any variation in the pulse shape. Therefore, we ascribe this effect to the space-time coupling induced by a misalignment of the pulse shaper.

Our results demonstrate that the pulse chirp variation reaching 100-200 $fs^2$ can be present even in the cases, where a single FROG trace of a 20 fs pulse can be reliably reproduced with a realistic pulse shape. In such cases, the D-FROG method provides a way to identify and quantify the variation level and, for instance, to identify misalignment of the pulse shaper.

## Funding

The authors gratefully acknowledge the financial support from The Czech Academy of Sciences (ERC-CZ/AV-B, Project Random-phase Ultrafast Spectroscopy; ERC100431901), and the Ministry of Education, Youth and Sports ("Partnership for Excellence in Superprecise Optics," Reg. No. CZ.02.1.01/0.0/0.0/16_026/0008390).

## Disclosures

The authors declare no conflicts of interest.